# Coupled thermodynamic-dynamic semi-analytical model of Free Piston Stirling engines

*Corresponding author*

Fabien Formosa, Laboratoire SYMME - Université de Savoie

BP 80439- 74944 ANNECY LE VIEUX CEDEX - FRANCE

Tel: +33 4 50 09 65 08

Fax: +33 4 50 09 65 43

Mail: fabien.formosa@univ-savoie.fr

*Author*

F. Formosa

*Affiliation*

Laboratoire SYMME, Université de Savoie, Annecy le Vieux, France

*Research highlights*

> The free piston Stirling behaviour relies on its thermal and dynamic features.

> A global semi-analytical model for preliminary design is developed.

> The model compared with NASA-RE1000 experimental data shows good correlations.

*Abstract*

The study of free piston Stirling engine (FPSE) requires both accurate thermodynamic and dynamic modelling to predict its performances. The steady state behaviour of the engine partly relies on non linear dissipative phenomena such as pressure drop loss within heat exchangers which is dependant on the temperature within the associated components. An analytical thermodynamic model which encompasses the effectiveness and the flaws of the heat exchangers and the regenerator has been previously developed and validated. A semi-analytical dynamic model of FPSE is developed and presented in this paper. The thermodynamic model is used to define the thermal variables that are used in the dynamic model which evaluates the kinematic results. Thus, a coupled iterative strategy has been used to perform a global simulation. The global modelling approach has been validated using the experimental data available from the NASA RE-1000 Stirling engine prototype. The resulting coupled thermodynamic-dynamic model using a standardized description of the engine allows efficient and realistic preliminary design of FPSE.

## Nomenclature

| *Symbol* | **Definition** |
|---|---|
| $A$ | Area [m$^2$] |
| $d$ | Intrinsic damping |
| $d_h$ | Hydraulic diameter [m] |
| $d_w$ | Fiber diameter of the matrix regenerator |
| $e$ | Regenerator efficiency [-] |
| $f$ | Frequency [Hz] |
| $f_F$ | Friction factor [-] |
| $h$ | Convection coefficient |
| $M$ | Total mass of the working fluid [kg] |
| $m$ | Mass of a sliding element [k] |
| $P$ | Power [W] |
| $p$ | Pressure [N m$^{-2}$] |
| $L_x$ | Length of exchanger [m] |
| $n_{Tx}$ | Number of tubes [-] |
| $r$ | Working fluid ideal gas constant |
| $r_h$ | Hydraulic radius of exchanger ($d_{hi}$ = 4 $r_{hi}$) |
| $s$ | Moving element stroke [m] |
| *T* | Temperature [°K] |
| *V* | Volume [m$^3$] |
| *Vsw* | Swept volume |
| *x* | Moving part position [m] |

*Greek symbols*

| | |
|---|---|
| ¶ | Porosity of the matrix regenerator |
| $\mu$ | Dynamic viscosity [N m$^{-2}$ s] |
| $\omega$ | Pulsation [rad s$^{-1}$] |
| $\alpha$ | Swept volume phase angle [rad] |
| $\alpha_{xdxp}$ | Displacer – piston phase angle [rad] |
| $\eta$ | Efficiency |
| $\Delta p$ | Pressure loss [N m$^{-2}$] |
| $\kappa$ | Swept volume ratio |
| $\xi$ | Expansion chamber to heat source temperatures ratio |
| $\delta$ | Cooler to heater heat exchange coefficients ratio |
| $\tau$ | Expansion to compression chamber temperatures ratio |
| $\Gamma$ | Heat source to sink temperatures ratio |

*Subscripts*

| | |
|---|---|
| *bp* | Piston bounce space |
| *bd* | Displacer gas spring space |
| *C* | Compression chamber |
| *CC* | Compression chamber dead volume |
| *d* | Displacer |
| *dc* | Compression side of the displacer |
| *de* | Expansion side of the displacer |
| *diss* | Dissipative relative term |
| *E* | Expansion chamber |
| *ff* | Free flow |
| *HC* | Expansion chamber dead volume |
| *h* | Heater |
| *k* | Cooler |
| *p* | Piston |
| *R* | Regenerator |
| *th* | Thermal relative term |
| *w* | Wetted |

## *1. Introduction*

One of the last and promising evolutions of the Stirling engine is the free piston mechanical arrangement designed by W. Beale at the end of the $60^{st}$ [1]. The main advantages are known as, a simpler mechanical design, no lateral loads which reduces wear and a following extended lifetime compared with classical Stirling engines. Therefore, applications such as radioisotope generator for deep space mission [2, 3] are currently under development.

Beside its advantages, the optimization of FPSE is a difficult task. Indeed, moving elements are driven by both the working fluid and gas springs pressures. The strokes and phase angle are then set by the coupled effect of the dynamic and thermodynamic parameters. As the volumes of the chambers are modified with the displacements of the piston and displacer, so are the pressure and the pressure losses through the heat exchangers. Moreover, the coupling intensity is related to the temperatures of the fluid within the engine. Hence, thermodynamic parameters of the engine have to be defined prior to any dynamic model of FPSE.

Due to the complexity of this analysis, the isothermal assumption is usually adopted. Furthermore, temperatures within the engine are given as input parameters. Linearization methods are then used in the vicinity of an operating point to obtain an estimation of the performances of the engine [4-8]. As a result, the behaviour of FPSE can be evaluated. Those analyses are limited to starting behaviour and the steady state characteristics can not be accurately modelled with linear approaches.

Ulusoy in [9] has studied the behaviour of FPSE using perturbation methods. Averaging or multiple scales methods have been used in many studies for simplification purpose of the effects of nonlinear phenomena which occur in many problems related to engineering and science [10]. The centre manifold approach is the one which leads to a semi-analytical simplified model. It accounts for the local bifurcation behaviour in the neighbourhood of a fixed point of the nonlinear system [11]. The centre manifold approach reduces the number of equations of the original system in order to obtain a simplified system without loosing the dynamics of the original system as well as the effects of nonlinear terms [12]. These studies have outlined the drastic effect of dissipative and non-linear phenomena on the FPSE behaviour. An equivalent global evaluation of the losses has been used and a straight relation between pressure loss and the exchangers characteristics *i.e.* length, hydraulic radius and free flow area can not be established. Consequently, these approaches appear to be unsuitable for preliminary design purpose and restricted to post analysis of existing design with given temperatures distribution. Therefore, there is a need for an accurate global approach including a thermodynamic isothermal model which can be used in accordance with dynamical analysis of the FPSE for preliminary design purpose.

Pressure drop loss evaluation is based on flow rate through each heat exchanger and regenerator. As far as isothermal assumption can be made, Organ in [13] has suggested a method to evaluate mass flow rates within the engine. In addition to geometric characteristics, they are linked to the operating frequency, pistons strokes, phase angle and chambers temperatures.

Temperatures within the engine are related to the heat exchangers performances and flow rate eventually. Representative thermodynamic models can be obtained by given heat source and sink temperatures instead of chambers ones [14]. Consequently, associated dynamic-thermodynamic studies can be used to obtain realistic results for FPSE models.

In a first part, a standardized description of FPSE for the thermodynamic and dynamic approaches is given. Then, the main steps of the proposed thermodynamic approach are recalled.

The evaluation of the mass flow rate and the related Reynolds number for the exchangers is detailed. It appears to be the corner stone of the approach. Values for heat transfer coefficient as well as the friction factor are given by experimental correlations [15]. Thus, it is possible to establish a strong relation between thermodynamic-dynamic model and design variables. Then, we recall the basis of the Hopf bifurcation analysis used to asses the dynamic analysis. The results of this semi analytical analysis are presented. Therefore, dynamic parameters can be tuned to match the thermodynamic ones. Finally, a joint thermodynamic-dynamic modelling strategy is proposed. The model is validated by comparison with the RE-1000 experimental data [16].

## 2. Analysis

### 2.1. Standardized description of FPSE

For any thermodynamic analysis, expansion and compression swept volumes ($V_{swE}$, $V_{swC}$) as well as their phase angle ($\alpha$) are supposed to be known. These parameters are forced by mechanical kinematics in classical Stirling engines. Mean pressure ($p_{mean}$), operating frequency ($\omega$), upper and lower external temperatures ($T_H$, $T_L$) are set as control parameters. Depending on the heat exchangers characteristics and the given working fluid, chamber temperatures $T_E$ and $T_C$ are to be determined. Engine performances can be deduced eventually.

On the contrary, in the case of a dynamical analysis of FPSE, chamber temperatures ($T_E$, $T_C$) are given. Mean pressure ($p_{mean}$) as well as working fluid are set. The problem at stake is to determine the operating pulsation ($\omega$), amplitudes and phase angle of displacer and piston $s_d$, $s_p$ and $\alpha_{xdxp}$ respectively. Swept volumes $V_{swE}$, $V_{swC}$ as well as their phase angle $\alpha$ are deduced and the engine performances can be evaluated eventually.

The classical Schmidt analysis [17] and dynamical model of FPSE which can be found in the literature use different definitions. The proposed preliminary design method presupposes to run both thermodynamic and dynamic analyses. Thus, a common set of parameters is required aiming at a global approach.

Relations between swept volumes and strokes amplitudes and phase are to be specified. For thermodynamic analysis, we set:

$$V_E(t) = V_{swE}/2\,(1 + \cos(\omega t))$$
$$V_C(t) = V_{swC}/2\,(1 + \cos(\omega t - \alpha)) \tag{1}$$

As harmonic periodic displacements are obtained for FPSE, swept volumes can be alternatively given. Setting, $x_d = s_d \cos(\omega t)$ and $x_p = s_p \cos(\omega t + \alpha_{xdxp})$, instantaneous volumes of the expansion and compression chambers are:

$$V_E(t) = A_{de}\, x_d(t)$$
$$V_C(t) = A_p\, x_p(t) - A_{dc}\, x_d(t) \tag{2}$$

Comparing equations (1) and (2), one can deduce:

$V_{swE} = 2 A_{de} s_d$

$$V_{swC} = 2\sqrt{(A_p s_p \cos(\alpha_{xdxp}) + A_{dc} s_d)^2 + (A_p s_p \sin(\alpha_{xdxp}))^2}$$

$$\alpha = \pi + \arctan\left[\frac{A_p s_p \sin(\alpha_{xdxp})}{-A_{dc} s_d + A_p s_p \cos(\alpha_{xdxp})}\right]$$
(3)

The total swept volume denoted $V_{sw}$ can be expressed from $V_{swE}$, $V_{swC}$ and $\alpha$. The swept volume ratio $\kappa = V_{swC} / V_{swE}$ is usually defined. Thus, total swept volume is:

$$V_{sw} = V_{swE}\sqrt{1 + \kappa^2 + 2\kappa \cos(\alpha)}$$
(4)

From the dynamic analysis, using (3), we can establish:

$$\kappa = A_{de} s_d / \sqrt{(A_p s_p \cos(\alpha_{xdxp}) + A_{dc} s_d)^2 + (A_p s_p \sin(\alpha_{xdxp}))^2}$$

$$V_{sw} = 2 A_{de} s_d \sqrt{1 + \kappa^2 - 2\kappa \frac{1}{\sqrt{1 + A_p^2 s_p^2 \sin(\alpha_{xdxp})^2 / (A_p s_p \cos(\alpha_{xdxp}) - A_{dc} s_d)^2}}}$$
(5)

## 2.2 Thermodynamic analysis

A general analysis scheme and associated parameters for any Stirling engine are presented in Fig. 1. In this section, main results of the thermodynamic study developed in [14] are given. This analysis relies on the usual Schmidt analysis (framed scheme of Fig 1) which allows expressing the indicated power as a function of chamber temperatures. In addition, a thermal model is defined in order to take into account the heat exchanger effectiveness as well as the regenerator reheat loss (see Fig. 1). It is possible to express the thermal power of the engine and the operating point can be determined by the equality between the two expressions.

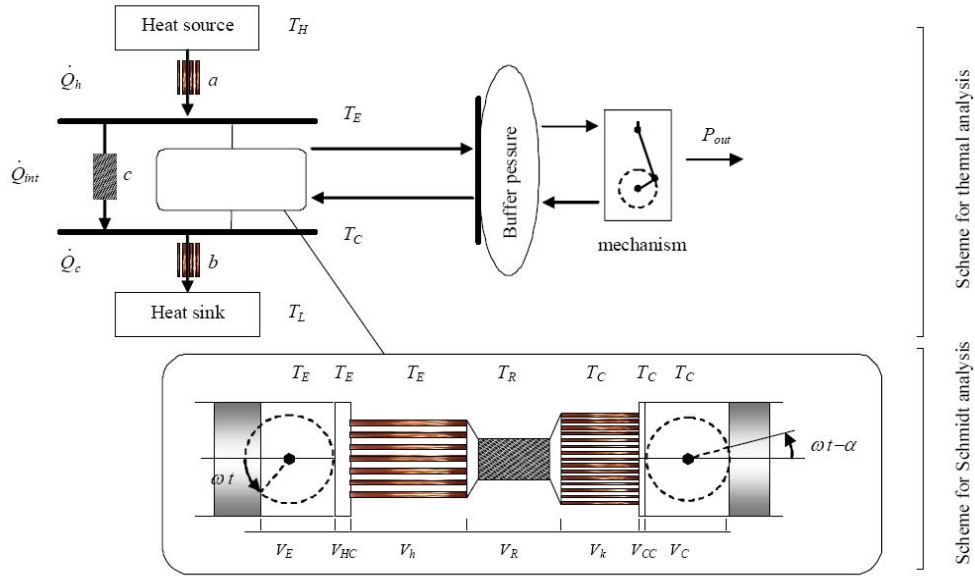

Fig. 1 Scheme for the thermodynamic analysis with heat exchangers and regenerator.

### 2.2.1. Results from the Schmidt analysis

The Schmidt analysis results expressed with both the thermodynamic and dynamic parameters are recalled hereafter.

The instantaneous pressure is:

$$p = p_{mean} \frac{\sqrt{1-\beta^2}}{1+\beta \cos(\omega t - \theta)} \tag{6}$$

Alternatively, it can be expressed as:

$$p = p_{mean} \frac{\sqrt{1-\beta^2}}{1 + \frac{1}{T_E}\frac{A_{dc}/\tau - A_{de}}{s} x_d - \frac{1}{T_C}\frac{A_p}{s} x_p} \tag{7}$$

wherein:

$$\beta = \frac{\sqrt{2\tau\kappa\cos(\alpha) + \tau^2 + \kappa^2}}{2\nu + \kappa + \tau}$$

$$\tan(\theta) = \frac{\kappa\sin(\alpha)}{\kappa\cos(\alpha) + \tau} \tag{8}$$

$$\tau = T_C / T_E$$

$$\nu = \frac{V_{CC}}{V_{swE}} + \frac{V_R}{V_{swE}} \frac{T_C}{T_R} + \frac{V_{HC}}{V_{swE}} \tau$$

$$\kappa = V_{swC} / V_{swE}$$

Consequently, the main characteristic parameters of the engine can be given by the following analytic expressions:

$$p_{mean} = \frac{M\,r}{s} \frac{1}{\sqrt{1-\beta^2}}$$

$$\frac{1}{s} = 2 \frac{T_E}{V_{swE}} \tau \frac{1}{2\nu + \kappa + \tau}$$

(9)

In which $M$ is the total mass of gas within the working spaces and $r$ the working fluid ideal gas constant.

The indicated power is expressed as:

$$P_i = p_{mean} \frac{\omega}{2\pi} \frac{V_{swE}}{2\beta} \frac{\kappa(\tau-1)\sin(\alpha)}{\sqrt{\tau^2 + \kappa^2 + 2\kappa\tau\cos(\alpha)}} (1 - \sqrt{1-\beta^2})$$

(10)

The ideal thermodynamic efficiency is the Carnot efficiency:

$$\eta_i = 1 - T_C / T_E$$ (11)

**2.2.2. Thermal approach**

- Thermal power

The cycle average power turns out to be given in an alternative way. According to the power balance of the machine as described as in Fig. 1, thermal power can expressed as:

$$P_{th} = h_h A_{wh} T_H (1 + \delta \Gamma - \xi - \xi \delta \tau)$$ (12)

In which $\xi = T_E / T_H$ is the temperature ratio between the heat source and the highest temperature of the engine and $\Gamma = T_L / T_H$.

From a simple convection model and given wetted area of the cooler and the heater ($A_{wk}$ and $A_{wh}$ respectively), $\delta$ can be expressed as:

$$\delta = \frac{h_k A_{wk}}{h_h A_{wh}} \qquad (13)$$

In which $h_h$ and $h_k$ are the convection heat transfers.

- **Thermal efficiency**

The thermal efficiency of the engine is defined by the ratio of the available power by the added thermal power:

$$\eta_{th} = \frac{1 + \delta\Gamma - \xi - \xi\tau\delta}{1 - \xi + (c_{cond} + (1-e)\dot{M}_R C_v)/(h_h A_{wh})\xi(1-\tau)} \qquad (14)$$

wherein $\dot{M}_R$ is the fluid mass rate inside the regenerator, $e$ is the regenerator effectiveness, and $C_v$ the specific heat for isochoric evolution.

- **Thermodynamic conditions**

From the second law of thermodynamics and an engine operation conditions, we set two constraint equations:

$$\eta_{th} \leq 1 - \tau$$
$$P_{th} \geq 0 \qquad (15)$$

Therefore, the permissible values of $\xi$ must be situated within hatched domain shown on Fig. 2.

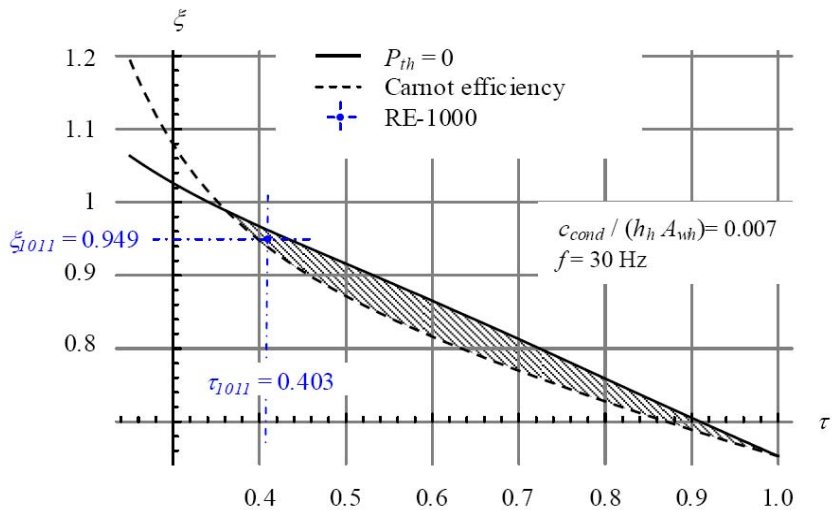

Fig. 2. Permissible area for $\xi$ for operating frequency of 30 Hz and mean pressure

As an example, result from experiment run #1011 of the RE-1000 [16] (see Table 1) is presented in Fig. 2. It is clearly located in the hatched region close to the Carnot efficiency limit.

As $\xi$ ratio must be defined as a single value, we choose here an optimal case. The inequalities Eq. 15 can be switched to equality. Consequently, the optimal ratio $\xi_{optim}$ between the heat source temperature and the temperature of the expansion chamber of the engine can be formally written as:

$$\xi_{optim} = \frac{\delta\Gamma + \tau}{(\delta+1)\tau - (c_{cond} + (1-e)\dot{M}_R C_v)/(h_h A_{wh})(1-\tau)^2} \qquad (16)$$

Finally, the operating point can be obtained matching the values of the indicated and thermal power given by (10) and (12) respectively. Figure 3 plots curves of temperature ratios $\tau$ and $\xi_{optim}$ for various operating frequencies $f = 2\pi/\omega$.

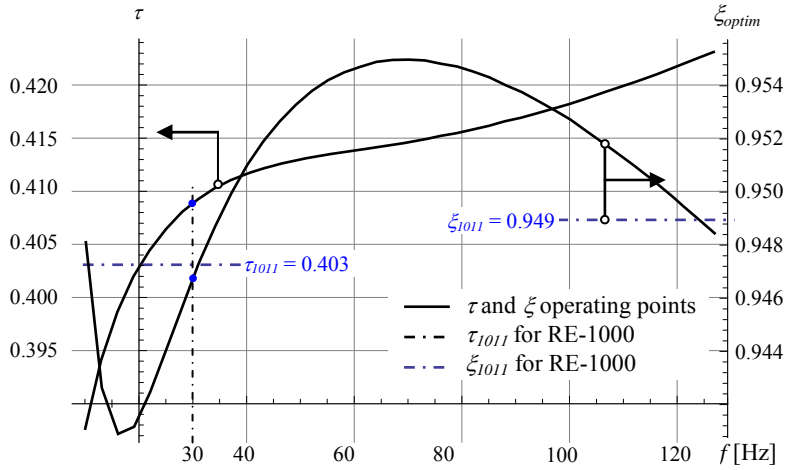

Fig. 3. Evolution of the operating point for various frequencies.

A discrepancy of 1.8% for the $\tau$ temperature ratio and 0.23% for the $\zeta$ temperature ratio can be evaluated.

Though, the temperature ratio can be an optimization criteria because it is related to the efficiency, both power and efficiency have to be considered. Figure 4 plots the power-efficiency curve for different frequencies. The experimental operating frequency of 30Hz of the RE-1000 run #1011 reference can be clearly identified as the optimal efficiency peak.

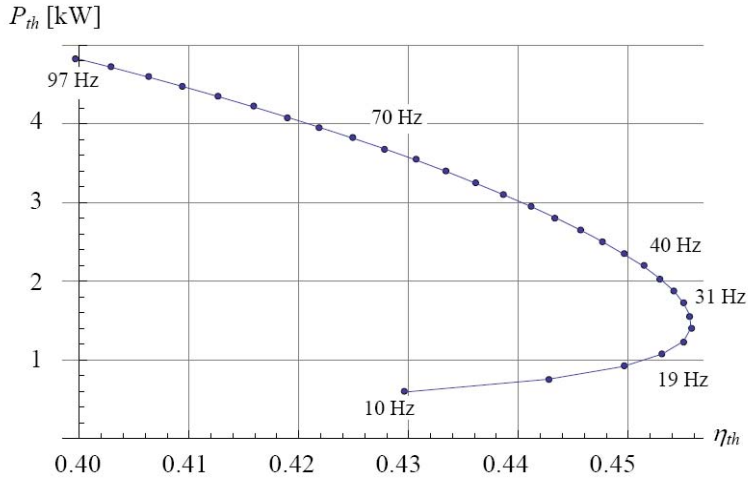

Fig. 4. Evolution of the Power-efficiency curve for various frequencies.

## 2.3. Analysis of the heat exchangers

The heat exchanger analysis appears to be a crucial facet of the engine model. Stirling engine performances rely on the heat transfer effectiveness. Moreover, the fundamental role of the regenerator itself leads to specific studies [18-19]. The dynamic aspect of the exchangers with respect to non linear dissipative phenomena has to be taken into account in a dynamic modelling of FPSE as a key effect as well.

For the thermal and dynamic points of view, usual experimental correlations can be used. Thus, the Colburn Factor is related to the heat transfer effectiveness whereas the friction factor allows pressure drop evaluation. These correlations are based on the Reynolds number which has to be determined for each of the exchanger. This evaluation is detailed hereafter.

### 2.3.1. Flow rate and Reynolds evaluation

The mass flow rate can be estimated from de Schmidt analysis as detailed in Organ work [13] dedicated to the study of the regenerator problem. Following the generic schematic of a Stirling engine (see framed scheme in Fig. 1), the mass of gas $M_x$ occupying a region between expansion piston face and any significant section of the engine (e.g. heat exchanger opposite end face) can be express as:

$$M_x = \frac{p}{r}\frac{1}{T_C}\tau(V_{swE}/2(1+\cos(\omega t))+V_h) \qquad (17)$$

Whereas the total mass of gas in the engine can be evaluated as:

$$M = \frac{p\,s}{r}(1 + \beta \cos(\omega t - \theta)) \tag{18}$$

Consequently, the mass fraction ratio results in a simple expression:

$$\frac{M_x}{M} = \frac{1}{s\,T_C} \frac{A + B\cos(\omega t)}{1 + C\cos(\omega t) + D\sin(\omega t)} \tag{19}$$

As an example the mass held between the displacer and the regenerator face of the heater:

$$\begin{aligned} A &= \tau(V_{swE}/2 + V_{HC} + V_h) \\ B &= \tau V_{swE}/2 \\ C &= \beta \cos\theta \\ D &= -\beta \sin\theta \end{aligned} \tag{20}$$

Therefore, it is possible to give the amount of fluid in the different chambers with respect to the position of the piston and displacer. Figure 5 plots the mass of fluid within the different parts of the engine. If there is no volume left in the compression chamber ($V_C = 0$), the whole mass of fluid is contained within the remaining spaces which is the case for the circle of the upper dashed line of Fig.5. When the schematic displacer reaches its upper position, there is no mass of fluid in the expansion space ($V_E = 0$) which is the case for the circle of the lower dashed line.

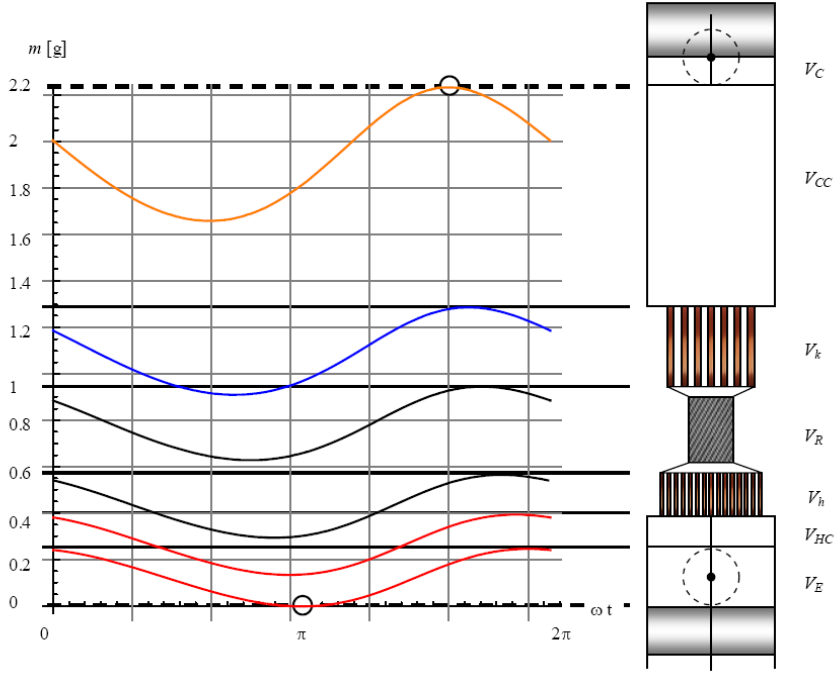

Fig. 5. Evolution of the mass of fluid within the different chambers of the RE-1000.

Note that for the RE-1000 the compression chamber dead volume represents about twice the cooler volume. The schematic on the right of Fig.5 is the attempted representation of the engine volumes.

From the evaluation of the masses, it is possible to give a mean value of mass flow rate for each of the exchanger and the regenerator. The regenerator case is detailed hereafter.

The interactive mass for the regenerator $M_{Rint}$ is defined by the maximum amount of the fluid mass between the displacer face and the compression output of the regenerator and the minimum value between the displacer and the regenerator expansion side. Theses masses are underlined in Fig. 6 which represents the evolution of $M_R$ and $M_h$, in spaces $\{V_E+V_{HC}+V_h\}$ and $\{V_E+V_{HC}+V_h+V_R\}$ respectively. The horizontal dashed line gives the value of $M_{R(int)}$. The resident mass of the regenerator is approximated by its mean value as (dotted-dashed line in Fig. 6):

$$M_{R(mean)} = \frac{p_{mean}}{r} \frac{1}{T_E} \frac{\log(1/\tau)}{1-\tau} V_R \qquad (21)$$

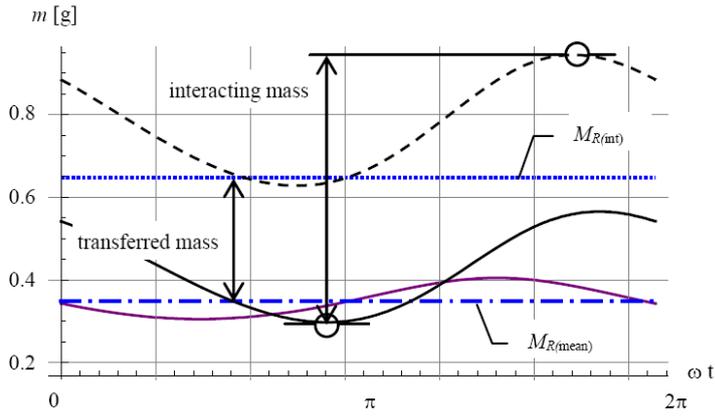

Fig. 6. Evolution of the mass of fluid within the regenerator.

Finally, the mean mass flow rate through regenerator is:

$$\dot{M}_R = \frac{\omega}{\pi}(M_{R(\text{int})} - M_{R(mean)}) \qquad (22)$$

Which enables to evaluate a Reynolds number as:

$$\text{Re}_R = \frac{\dot{M}_R}{\mu}\frac{4\,rh_R}{\text{Aff}_R} \qquad (23)$$

Figure 7 plots the Reynolds number for the regenerator as a function of the temperature ratio $\tau$ from the RE-100 parameters for different operating frequencies. Due to the very small hydraulic radius, a laminar flow is obtained for the regenerator. For the heater and the cooler the usually reported turbulent behaviour of mean flow is predicted as well.

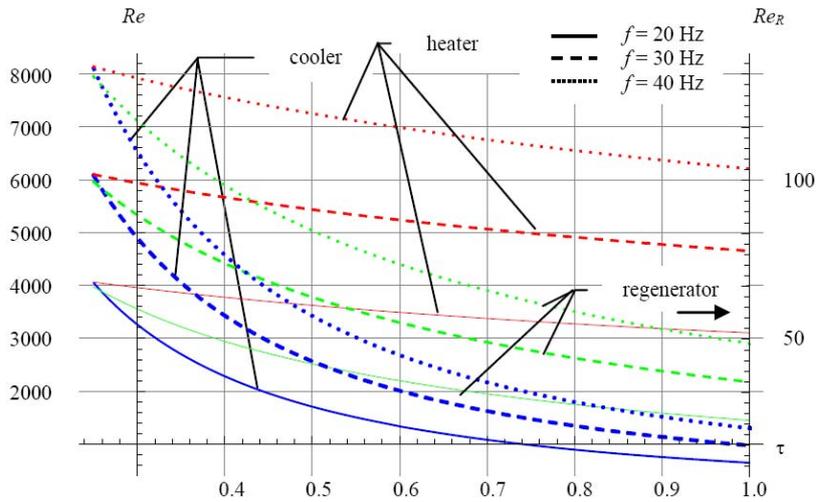

Fig. 7. Evolution of the Reynolds numbers for heat exchangers and regenerator with respect to temperature ratio $\tau$ for three different operating frequencies.

Once the Reynolds numbers are evaluated for each exchanger, it is possible to use the experimental correlations as a simple way to deduce both their heat transfers and friction behaviours. As a keypoint for the dynamic analysis we focus here on the later.

### 2.3.2. Pressure drop coefficient

Different from the work of De Monte [4,5], evaluations of individual pressure drops related to each of the heat exchangers are seeking here. The friction factor can be estimated [15] from Reynolds numbers. Thus, it is possible to give an estimation of the pressure drop in heater, cooler and regenerator denoted $\Delta p_h$, $\Delta p_k$ and $\Delta p_R$ respectively, with respect to their geometrical parameters but also on the operating parameters $V_{sw}$, $\kappa$, $\alpha$ and $\omega$. Pressure drops can be calculated using the friction factor coefficient related to the Reynolds number. As a consequence the impact of the geometric parameters on the engine losses and its operation and performances can be studied.

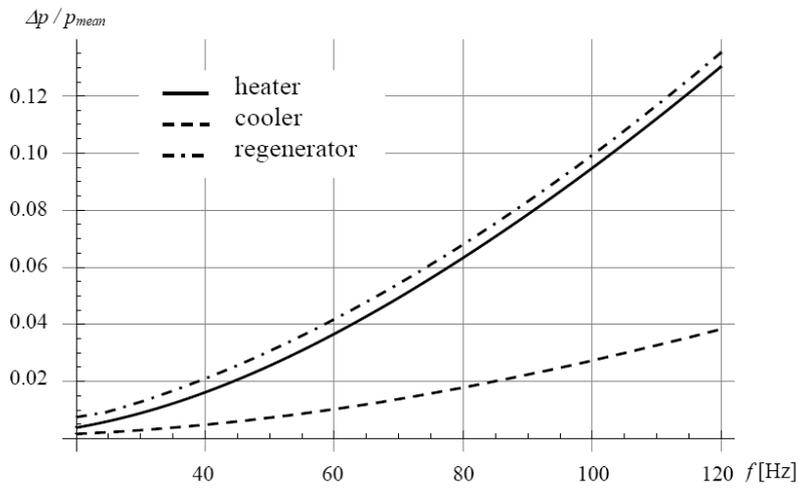

Fig. 8. Evolution of the relative pressure drop for various operating frequencies.

Figure 8 plots the relative pressure drop with respect to the operating frequency for each heat exchanger and regenerator.

## 2.4. Dynamic analysis

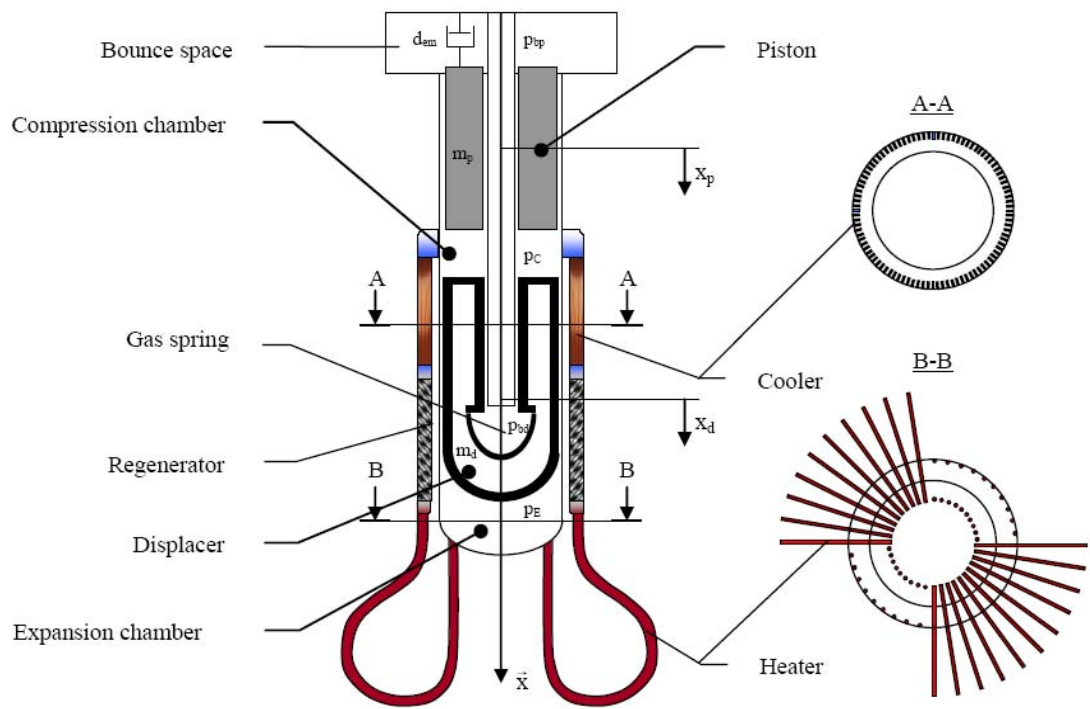

Fig. 9. Scheme for the dynamic analysis of FPSE.

### 2.4.1. Governing equations

Assuming that the space coordinates $x_p$ and $x_d$ are relative to the midpoint of the stroke for each moving part (piston, displacer). Besides, the case movement is neglected. The equations that describe the dynamic behaviour of an FPSE can be written for each of the moving part such that:

$$m_d \ddot{x}_d + d_d \dot{x}_d = (A_{bd} \, p_{bd} - A_{de} \, p_E + A_{dc} \, p_C)$$
$$m_p \ddot{x}_p + (d_p + d_{em}) \dot{x}_p = (A_p \, p_{bp} - A_p \, p_C) \tag{24}$$

with the following initial conditions:

$$x_p|_{t=0} = 0, \; \dot{x}_p|_{t=0} = 0 \, ; \, x_d|_{t=0} = 0 \, ; \, \dot{x}_d|_{t=0} = 0 \tag{25}$$

The pressures $p_E$, $p_C$ are the pressure in expansion and compression spaces as defined in Fig. 9. $p_{bd}$, $p_{bp}$ are pressures in the gas spring and bounce space of the engine. The effect of the mechanical dissipative phenomena is taken into account by means of the $d_p$ and $d_d$ damping coefficients.

This model would be validated using the data from the RE-1000 experiments. A dashpot load is used as a control parameter that determined the piston, displacer and pressure amplitudes and the engine frequency. As a consequence, we choose to define a simple control parameter as a viscous damping associated to the piston and denoted $d_{load} = d_p + d_{em}$.

In equation (24) $x_p$ ($x_d$) is the position of the piston (displacer) with respect to its rest equilibrium position. $\dot{x}_p$ ($\dot{x}_d$) is the derivative of $x_p$ ($x_d$) with respect to time.

The instantaneous pressure $p$ within the chambers is analytically expressed as in (7) which is more suitable than (6). Besides, pressures of the chambers are linked by the pressure losses which occur in the heat exchangers as well as in the regenerator. Thus, we set: $p_E = p + (\Delta p_h + \Delta p_R + \Delta p_k)$ and $p_C = p$.

Pressure in the gas spring and the buffer space may be evaluated by assuming the ideal gas relation for an adiabatic process:

$$p_g = p_{mean}\sqrt{1-\beta^2}\left(\frac{V_{s0}}{V_s}\right)^{\gamma} \quad (26)$$

Therefore, the general form of equation (24) can be given in the following way:

$$\begin{bmatrix}\dot{x}_d \\ \ddot{x}_d \\ \dot{x}_p \\ \ddot{x}_p\end{bmatrix} = \begin{bmatrix} 0 & 1 & 0 & 0 \\ -S_{dd} & -D_{dd} & -S_{dp} & -D_{dp} \\ 0 & 0 & 0 & 1 \\ -S_{pd} & -D_{pd} & -S_{pp} & -D_{pp} \end{bmatrix}\begin{bmatrix}x_d \\ \dot{x}_d \\ x_p \\ \dot{x}_p\end{bmatrix} + \begin{bmatrix} 0 \\ fNL_d \\ 0 \\ fNL_p \end{bmatrix} \quad (27)$$

with initial conditions given by equation (25)

The expressions of the pressure losses $\Delta p_e$, $\Delta p_k$ and $\Delta p_R$ are nonlinear relations in which the classical Reynolds number is the main variable. Pressure terms are approximated evaluating Taylor's development so is the pressure expression (7). The nonlinear stiffness and dissipative terms are grouped in $fNL_d$ and $fNL_p$.

### 2.4.2. Solution procedure

Equation (27) can also be seen as two nonlinear oscillators coupled by mechanical forces. A classical supercritical Hopf bifurcation defines the steady oscillations in such coupled oscillators. At the fixed point also called the bifurcation point denoted $x_0$, a couple of complex conjugated eigenvalues associated to the system crosses the imaginary axes. Hence, this point becomes unstable and because of the nonlinearities of the system a stable limit cycle can occurs in its vicinity as can be seen in Fig. 10. The effect of the load of the system represented by the damping coefficient $d_p$ is reflected in Fig. 10 in which its increasing leads to the stabilization of the system. Thus, the load may have to be controlled during the starting phase.

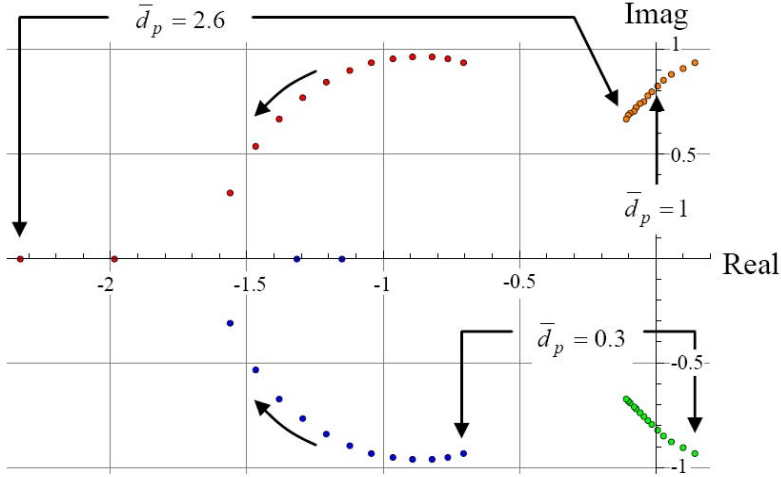

Fig. 10. Evolution of real and imaginary parts of the system's poles with respect to the reduced damping coefficient $\bar{d}_p = d_{load} / d_{load\,crit}$.

- **Unstable fixed point**

The fixed point $[x_d\,;\dot{x}_d\,;x_p\,;\dot{x}_p] = [0;0;0;0]$ is obtained by solving the non-linear static equations. The evaluation of the eigenvalues of the constant Jacobian matrix at the fixed point $q_0$ allows the stability analysis. They are defined as roots of the fourth-degree characteristic polynomial such as:

$$r^4 + \alpha\,r^3 + \beta\,r^2 + \gamma\,r + \delta = 0 \qquad (28)$$

in which :

$$\alpha = D_{pp} + D_{dd}$$

$$\beta = D_{pp} D_{dd} - D_{pd} D_{dp} + S_{dd} + S_{pp}$$

$$\gamma = -D_{dp} S_{pd} - D_{pd} S_{dp} + D_{dd} S_{pp} + D_{pp} S_{dd} \qquad (29)$$

$$\delta = S_{pp} S_{dd} - S_{pd} S_{dp}$$

Depending on the operating parameters, the only situation for which oscillations can occur is: two complex conjugate roots with positive real parts and two complex conjugate roots with negative real parts. Thenceforth, a further analysis is required to asses the steadiness of these oscillations.

The roots of (28) evaluated at the fixed point can be expressed as two complex conjugates expressions:

$$\lambda_{Jc1} = \alpha_1 - j\,\omega_1 \;;\; \lambda_{Jc2} = \alpha_1 + j\,\omega_1 \tag{30}$$

$$\lambda_{Js1} = \alpha_2 - j\,\omega_2 \;;\; \lambda_{Js2} = \alpha_2 + j\,\omega_2 \tag{31}$$

in which j is the complex number such as $j = \sqrt{-1}$ and $\alpha_1$ is any positive real number.

- Centre manifold approach

The previous linear analysis allows defining the condition necessary to the engine operation but can not represent the behaviour of the engine after the starting point. Therefore, another theory has to be used to study the post bifurcation behaviour. This section summarizes the application of the centre manifold method [12] to study the system defined by (27). The stability of the centre manifold is equivalent to the stability of the original system in the vicinity of the fixed point. Again, the bifurcation appears when one or several eigenvalues cross the imaginary axis in the complex plane with the variation of a control parameter (see Fig. 10).

At the Hopf bifurcation point, $\text{Re}(\lambda_{Jc1}) = 0$ and Eq. (32) can be written in the canonical form:

$$\begin{aligned}\dot{\boldsymbol{v}}_c &= \mathbf{J}_c\,\boldsymbol{v}_c + \mathbf{G}_2(\boldsymbol{v}_c, \boldsymbol{v}_s) + \mathbf{G}_3(\boldsymbol{v}_c, \boldsymbol{v}_s) \\ \dot{\boldsymbol{v}}_s &= \mathbf{J}_s\,\boldsymbol{v}_s + \mathbf{H}_2(\boldsymbol{v}_c, \boldsymbol{v}_s) + \mathbf{H}_3(\boldsymbol{v}_c, \boldsymbol{v}_s)\end{aligned} \tag{32}$$

wherein $\mathbf{J}_c$ et $\mathbf{J}_s$ have eigenvalues such as $\lambda_{Jc} = \alpha_1 +/- j\,\omega_1$ and $\lambda_{Js} = \alpha_2 +/- j\,\omega_2$. $\mathbf{G}_2$, $\mathbf{G}_3$, $\mathbf{H}_2$ and $\mathbf{H}_3$ are vector for which the two components are polynomials of degree 2 and 3 in the components of vectors $\boldsymbol{v}_c$ and $\boldsymbol{v}_s$ namely $v_{c1}$, $v_{c2}$ and $v_{s1}$, $v_{s2}$.

For the FPSE case, it may be assumed that $\alpha_2$ is negative. $\boldsymbol{v}_s$ defines a stable subspace which is the linear approximation to the stable manifold. When we choose an initial condition on this stable manifold sufficiently close to the fixed point the solution curve will go toward the fixed point. In accordance to the parameters, an unstable subspace defined by the eigenvalues of matrix **A** such as their real part is zero can exist. The centre manifold theorem allows to represent locally the centre manifold as { $[\boldsymbol{v}_c, \boldsymbol{v}_s]$ such that $\boldsymbol{v}_s = \mathbf{h}(\boldsymbol{v}_c)$, $\mathbf{h}(0) = \mathbf{0}$, $D\mathbf{h}(0) = \mathbf{0}$ } and consequently reduce the four dimensional initial problem to a two dimensional one.

Substituting into the second line of equation (32) one obtains:

$$\dot{\boldsymbol{v}}_c = \mathbf{J}_c\,\boldsymbol{v}_c + \mathbf{G}_2(\boldsymbol{v}_c, \boldsymbol{v}_s) + \mathbf{G}_3(\boldsymbol{v}_c, \boldsymbol{v}_s) \tag{33}$$

$$D_{\boldsymbol{v}_c}(\mathbf{h}(\boldsymbol{v}_c))\,\dot{\boldsymbol{v}}_c = \mathbf{J}_s\,\mathbf{h}(\boldsymbol{v}_c) + \mathbf{H}_2(\boldsymbol{v}_c, \mathbf{h}(\boldsymbol{v}_c)) + \mathbf{H}_3(\boldsymbol{v}_c, \mathbf{h}(\boldsymbol{v}_c)) \tag{34}$$

Thus, it is possible to define an approximate solution **h** using a power expansion without constant and linear terms:

$$\mathbf{h}(\mathbf{v}_c) = \sum_{p=i+j=2}^{m} \sum_{i=0}^{p} \sum_{j=0}^{p} a_{ij}\, v_{c1}{}^{i}\, v_{c2}{}^{j} \tag{35}$$

Replacing $\dot{v}_c$, the complex coefficients $a_{ij}$ can be obtained. As a result, the dynamic behaviour of the system is determined by the following reduced system:

$$\dot{v}_c = \mathbf{J}_c\, \mathbf{v}_c + \mathbf{G}_2(\mathbf{v}_c,\, \mathbf{h}(\mathbf{v}_c)) + \mathbf{G}_3(\mathbf{v}_c,\, \mathbf{h}(\mathbf{v}_c)) \tag{36}$$

- Normal form analysis

The normal form analysis aims at a transformation of a system of nonlinear equations through a sequence of nonlinear-near identity transformations to eliminate as many nonlinear terms as possible. Those which cannot be removed are called the secular or the resonant terms. This simplest form of the equations is called the "normal form". The analysis of the dynamics of the normal forms reveals a qualitative picture of the flows of each bifurcation type.

For the sake of clarity, equation (36) is written again as below:

$$\dot{v}_c = \alpha_1 v_c - \omega_1 \bar{v}_c + f(v_c, \bar{v}_c)$$
$$\dot{\bar{v}}_c = \omega_1 v_c + \alpha_1 \bar{v}_c + g(v_c, \bar{v}_c) \tag{37}$$

in which with $\bar{v}_c$ is the complex conjugate of $v_c$.

As set of equations (37) are a pair of complex conjugate equations; one single equation needs to be studied eventually. A Hopf bifurcation is identified here and the result of these successive transforms is known [12].

$$\dot{v}_c = \alpha_1 v_c - \omega_1 \bar{v}_c + (C_1 v_c - C_2 \bar{v}_c)(v_c^2 + \bar{v}_c^2) \tag{38}$$

with:

$$16\, C_1 = (f_{xxx} + f_{xyy} + g_{xxy} + g_{yyy}) + 1/\omega_1\, [\, f_{xy}\,(f_{xx} + f_{yy}) - g_{xy}\,(g_{xx} + g_{yy}) - f_{xx}g_{xx} + f_{yy}g_{yy}\,] \tag{39}$$

$$16\, C_2 = (f_{xxy} + f_{yyy} - g_{xyy} - g_{xxx}) + 1/\omega_1\, [\, 2\,(f_{xx}^2 + f_{xy}^2) + 5\,(f_{yy}^2 + 5g_{xx}^2) + 5f_{xx}\,(f_{yy} - g_{xy}) \tag{40}$$

$$+ 2 g_{xy}^2 - f_{yy}g_{xy} + 5g_{xx}g_{yy} + 2g_{yy}^2 - f_{xy}(g_{xx} + 5g_{yy})\,]$$

In which $f_x$ ($g_x$) is the partial derivative to the first order of the function $f(g)$ with respect to the variable $x$ and $f_{xx}$ ($g_{xx}$) is the partial derivative to the second order of a function $f(g)$ with respect to the variable $x$. Using polar form ($v_c = r\cos\theta$ and $\bar{v}_c = r\sin\theta$), we finally obtain:

$$\dot{r} = \alpha_1 r + C_1 r^3$$
$$\dot{\theta} = \omega_1 r + C_2 r^2 \tag{41}$$

The amplitude of a limit cycle can be assed by the first of the previous equation. The non trivial solution establishes the existence condition for the limit cycle as: $\alpha_1/C_1 < 0$. The stability of each fixed point can be studied by the sign of the Jacobian $J(r)$. From (41) $J(r) = \alpha_1 + 3 C_1 r^2$, therefore $J(r_{f1}) = \alpha_1$ and $J(r_{f2}) = -2\alpha_1$. Thus, if $\alpha_1$ is lower than zero and a greater than zero, the only stable fixed point is $r_{f1} = 0$ and if $\alpha_1 > 0$ and $a < 0$, the unique stable fixed point is $r_{f2}$ and a limit cycle occurs.

Figure 11 plots the results of the equations (27) using a direct numerical solution and the semi analytical approach previously described. As a comparison, the displacement of the piston and displacer from RE-1000 experimental data #1011 are plotted.

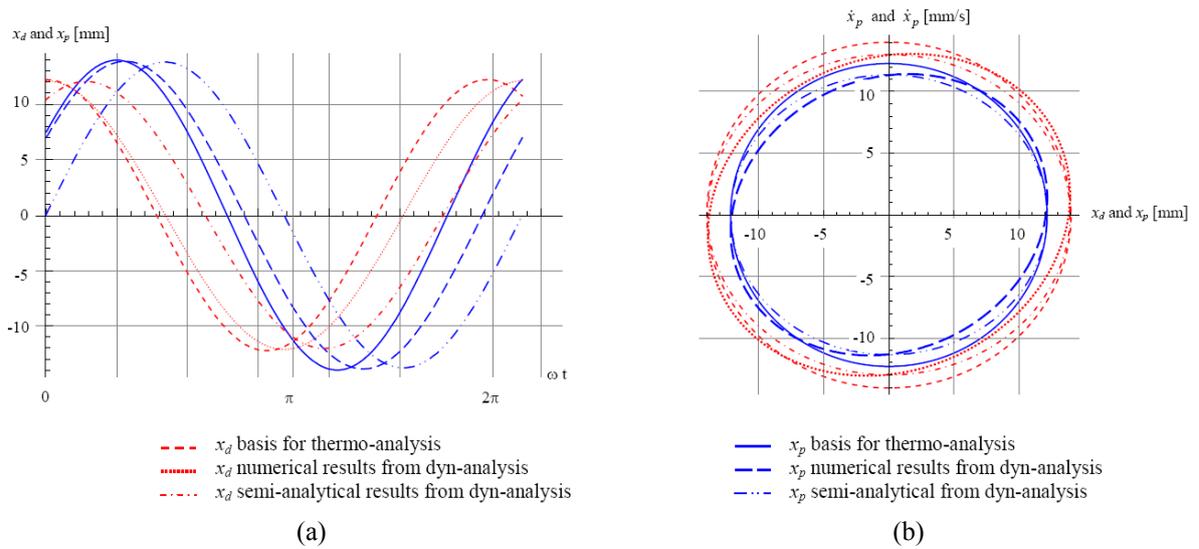

(a)

- - - $x_d$ basis for thermo-analysis
........ $x_d$ numerical results from dyn-analysis
-..-.. $x_d$ semi-analytical results from dyn-analysis

(b)

—— $x_p$ basis for thermo-analysis
- - - $x_p$ numerical results from dyn-analysis
-..-.. $x_p$ semi-analytical from dyn-analysis

Fig. 11. Piston and displacer movements.

Results for numerical and semi-analytical approaches match. The strokes are well determined (see Fig. 11 (b)) but the phase angle $\alpha_{xdxp}$ and the frequency show slight discrepancy. Comparisons details will be given in next section.

## 3. Model validation - RE-1000 comparison

In order to validate the proposed modelling strategy, thoroughly documented FPSE is needed. One of the available documented FPSE is the RE-1000 studied by the NASA during the 80's [16]. The purpose of the tests was to collect data over a wide range of operating conditions. Changes in the operating condition were accomplished by varying the working fluid, the mean pressure level, the heat source and sink temperatures. A dashpot load enabled to define the piston amplitude and displacer, pressure amplitudes and the engine frequency eventually. The description of one of the selected test configuration is given hereafter in Table 1. Some of the relevant variables (*e.g.* $V_{sw}$, $\alpha$, $K$) for these run are deduced from the raw data with equations (3-5).

| Heat source $T_H$ | [°C] | 600 | Cold source $T_C$ | [°C] | 25 |
|---|---|---|---|---|---|
| Phase angle $\alpha$ | [deg] | 106.44 | swept volume ratio $\kappa$ | [ - ] | 1.022 |
| Swept volume $V_{sw}$ | [cm$^3$] | 74.90 | Working fluid | | Helium |
| Mean pressure $p_{mean}$ | [MPa] | 7.034 | Overall efficiency $\eta_{tm}$ [%] | | 23.7 |
| Operating frequency $f$ | [Hz] | 30.1 | Output power [W] | | 955 |
| Wire mesh regenerator 1 | | | regenerator length | [mm] | 64.46 |
| wire diameter $d_w$ | [µm] | 88.9 | porosity of matrix $\P_v$ | [ - ] | **0.759** |
| regenerator hydraulic radius $r_{hR}$ [mm] | | 0.07 | material conductivity | [Wm$^{-1}$K$^{-1}$] | 16.3 |
| Displacer 1 | | | rod diameter | [mm] | 16.657 |
| displacer weight | [g] | 426 | Gas spring volume | [cm$^3$] | 31.79 |

Table 1. Parameter values of Run #1011.

### 3.1. Iterative evaluation strategy

Thermodynamic and dynamic results are obtained using a two steps iterative strategy. Figure 12 describes the general iterative scheme. The strategy used to couple the two models is first to obtain an evaluation of the chambers temperatures thanks to thermodynamic analysis using kinematic parameters from RE-1000 data.

Second, these values are input parameters for dynamic analysis for which the linear viscous damping coefficients $d_d$, $d_p$ and $d_{load}$ are used as calibration variables. They are modified to match the strokes of the piston and displacer as well as their phase angle in addition to the operating frequency. As defined by equations 3-5, swept volume, phase angle and swept volume ratio can be evaluated eventually.

After a first iteration for the thermodynamic plus dynamic analysis, the evaluated kinematic results are parameters for a next iteration of thermodynamic analysis. The resulting temperatures $T_E$ and $T_C$ are compared with the previous values. A relative difference less than $\varepsilon_T = 0.5\%$ ensures the convergence of the thermodynamic analysis. This convergence criterion is denoted $TV_{i+1} - TV_i \leq \varepsilon_T$ in Fig. 13.
The Reynolds values and the deduced pressure drop coefficients are strongly dependent on the kinematic variables as can be seen from (20). As a consequence, a second dynamic analysis is performed using the new thermodynamic results as parameters. New deduced operating values ($V_{sw}$, $\kappa$, $\alpha$ and $\omega$) are evaluated using equations (3-5). These kinematic variables are defined as $KV$. A convergence criterion is defined as: $KV_{i+1} - K_{Vi} \leq \varepsilon_T$. The maximal difference between successive results have to be less than $\varepsilon_K = 2\%$.

It is worthy of note that the pressure drops are not considered here as calibration parameters. Moreover, save for the very first iteration of all the comparison cases, the piston load is the single dynamic calibration parameter.

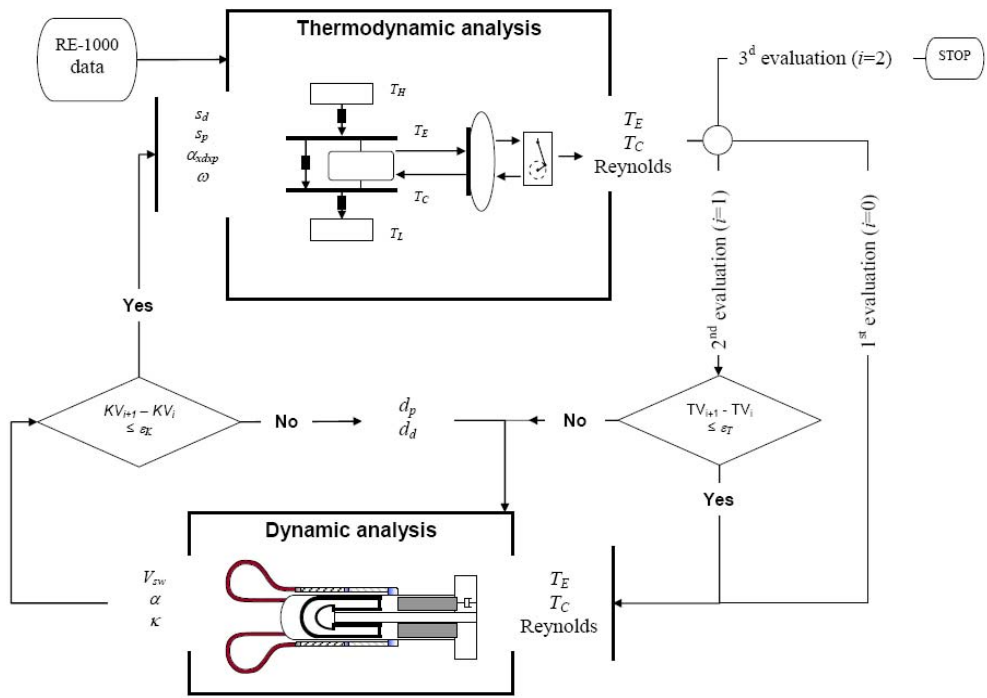

Figure 12 plots the iterative strategy used for the global analysis.

### 3.2. RE-1000 run #1011 comparison

As a first example, a detailed presentation is given for experimental run #1011.

- First iteration: Thermodynamic analysis

Input data and results from thermodynamic analysis first evaluation are given in Table 2. Bold figures represent the estimated values from the model whereas input data are given in gray background for information. For each calculated value, relative deviation with the corresponding experimental reference value is given in brackets. Temperature evaluations are in good agreement with experimental ones.

| Heat source $T_H$ | [°C] | 600 | Cold source $T_C$ | [°C] | 25 |
|---|---|---|---|---|---|
| **Expansion temperature $T_E$ [°C]** | | **554.4 (-0.1%)** | **Compression temperature $T_C$ [°C]** | | **66.3 (1.6%)** |
| Phase angle $\alpha$ | [deg] | 106.4 | **temperature ratio $\tau$** | [ - ] | **0.41 (-1.7%)** |
| Swept volume $V_{sw}$ | [cm³] | 74.90 | swept volume ratio $\kappa$ | [ - ] | 1.022 |
| Operating frequency $f$ | [Hz] | 30.1 | Mean pressure $p_{mean}$ | [MPa] | 7.03 |

Table 2. Results from thermodynamic analysis $i$=0.

- First iteration: Dynamic analysis

Input data and results are given in Table 3. Bold figures represent the estimated values from the model whereas input data from the thermodynamic results are given for information.

| Heat source $T_H$ | [°C] | 600 | Cold source $T_C$ | [°C] | 25 |
|---|---|---|---|---|---|
| Expansion temperature $T_E$ [°C] | | 560.8 | Compression temperature $T_C$ [°C] | | 66.3 |
| **Phase angle $\alpha$** | **[deg]** | **108.11 (-1.6%)** | temperature ratio $\tau$ | [ - ] | 0.41 |
| **Swept volume $V_{sw}$** | **[cm³]** | **74.95 (0.06%)** | **swept volume ratio $\kappa$** | **[ - ]** | **1.06 (-4.4%)** |
| **Operating frequency $f$** | **[Hz]** | **27.8 (7.7%)** | Mean pressure $p_{mean}$ | [MPa] | 7.03 |

Table 3 Results from dynamic analysis from first ($i$=0) thermodynamic analysis results parameters.

The output power is evaluated through the linear damping of the piston: Poutdyn = 1166 W. Compare to the RE-1000 output power Pout = 955 W the model gives a first quite good estimation (22 % discrepancy).

Figure 13 a) plots Clapeyron diagrams for the expansion and the compression chamber. The results from the dynamic and the thermodynamic first analysis are presented and appear to be very close to each other.

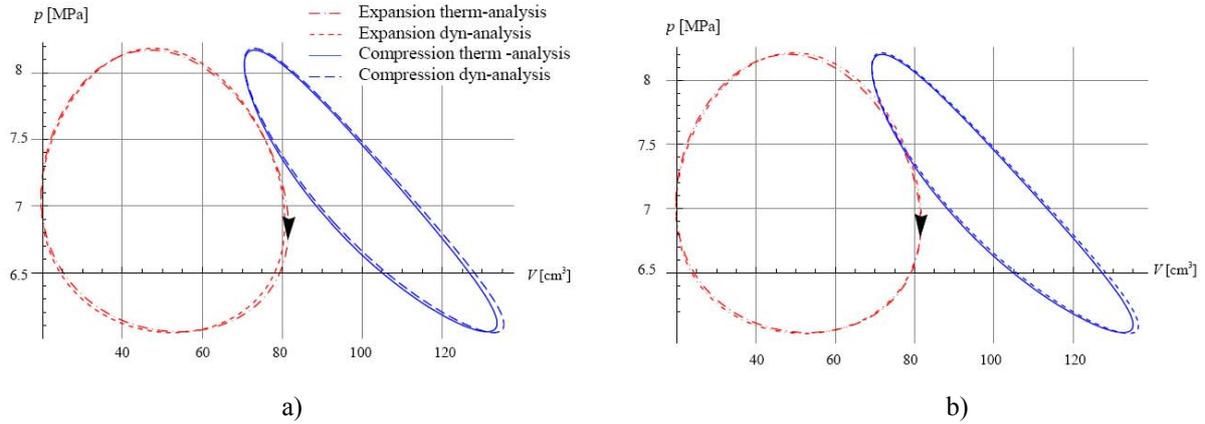

Fig. 13. Clapeyron diagrams for compression and expansion spaces. a) first iteration, b) second iteration

Though the convergence criteria is satisfied, new values for Reynolds number have to be evaluated again in order to take into account drift related to swept volume ratio and operating frequency. Modified pressure drops values have strong influence on the nonlinear characteristics of the system. As a consequence the dissipative load variable will be adapted and so do the output power.

- Second iteration: Thermodynamic analysis

For the second iteration, kinematic results from the dynamic analysis are set as input parameters. Bold figures of Table 4 represent the estimated values from the model whereas input data from the previous iteration are given for information. For each calculated value, the drift with the previous evaluation is also given.

| Heat source $T_H$ | [°C] | 600 | Cold source $T_C$ | [°C] | 25 |
|---|---|---|---|---|---|
| **Expansion temperature $T_E$ [°C]** | | **552.9 (0.18%)** | **Compression temperature $T_C$ [°C]** | | **65.5 (0.26%)** |
| Phase angle $\alpha$ | [deg] | 108.11 | **temperature ratio $\tau$** | **[ - ]** | **0.41 (0.08%)** |
| Swept volume $V_{sw}$ | [cm³] | 74.09 | swept volume ratio $\kappa$ | [ - ] | 1.06 |
| Operating frequency $f$ | [Hz] | 27.8 | Mean pressure $p_{mean}$ | [MPa] | 7.03 |

Table 4. Results from thermodynamic analysis ($i$=1).

Temperature drifts are small which ensures convergence ($\varepsilon_T$ < 0.5%).

- Second iteration: Dynamic analysis

| Heat source $T_H$ | [°C] | 600 | Cold source $T_C$ | [°C] | 25 |
|---|---|---|---|---|---|

| Expansion temperature $T_E$ [°C] | 552.9 | Compression temperature $T_C$ [°C] | 65.5 |
|---|---|---|---|
| Phase angle $\alpha$ [deg] | **110.3 (-2.0%)** | temperature ratio $\tau$ [ - ] | 0.41 |
| **Swept volume $V_{sw}$ [cm³]** | **72.60 (1.7%)** | **swept volume ratio $\kappa$ [ - ]** | **1.08 (-1.8%)** |
| **Operating frequency $f$ [Hz]** | **27.8 (-0.3%)** | Mean pressure $p_{mean}$ [MPa] | 7.03 |

Table 5. Results from dynamic analysis ($i$=1).

From the results of from the dynamic analysis given in Table 5, the output power evaluated through the linear damping of the piston: $P_{out}$ = 1104 W. Compare to the RE-1000 output power $P_{outexp}$ = 955 W the model gives a good estimation (15.6 % discrepancy).

It is worthy of note that the dissipated power at the displacer is predicted as about 114 W which is about 10% of the piston power. As discrepancies for kinematic variables show less than 2% difference with the first iteration the coupling of the approaches appears to be a stable process. After the second iteration, Clapeyron diagrams for the expansion and compression chambers show a good agreement in Fig. 13 b).

- Final comparison with run #1011 data

Table 6 gives the final results after the two steps procedure for the global analysis.

| Expansion temperature $T_E$ [°C] | 553.2 (-0.27%) | Compression temperature $T_C$ [°C] | 65.0 (1.17%) |
|---|---|---|---|
| Piston stroke [cm] | 2.72 (-2.6%) | Displacer stroke [cm] | 2.41 (-2%) |
| Displacer – piston phase angle [deg] | 62.2 (8.2%) | | |
| Phase angle $\alpha$ [deg] | 110.3 (3.64%) | | |
| Swept volume $V_{sw}$ [cm³] | 72.60 (-3.08%) | swept volume ratio $\kappa$ [ - ] | 1.08 (6.16%) |
| Operating frequency $f$ [Hz] | 27.9 (-7.47%) | Output power $P_{out}$ [kW] | 1.1 (15.60%) |
| Overall efficiency $\eta_{tm}$ [%] | 29.2 (23.3%) | Input thermal power[*] [kW] | 3777 (-6.46%) |
| | | Output thermal power [kW] | 2107 (-30.5%) |

Table 6. Final comparison for run #1011.

[*] For the experimental results, the electrical power input is used as an evaluation of the power input.

The proposed global analysis allows an estimation of the kinematic results close to 10%. Compare to linear dynamic analysis, the global approach gives better results and is able to evaluate thermodynamic as well as dynamic variables.

Table 7 gives comparisons of dynamical models from [6] with experimental results. For the chosen kinematic results, the proposed model appears to be closer than any of the approaches. Because the evaluated temperatures are also close to the experimental ones (see Table 6), performances of the engine can be well predicted eventually.

| Parameter | Urieli and Berchowitz. | Walker and Senft | Rogdakis case study 1 | Rogdakis case study 2 | Proposed global analysis |
|---|---|---|---|---|---|
| Operating frequency $f$ [%] | 10.7 | 9.7 | 0 | 0.3 | -7.5 |
| Displacer – piston phase angle [%] | 36.2 | 29.6 | 28.9 | 29.4 | 8.2 |
| Amplitude ratio [%] | -41.5 | –40.6 | -17.9 | -31.1 | -0.9 |

Table 7. Results from different models of RE-1000

## 3.2. Influence of piston stroke amplitude for displacer 1

Experimental procedure used in [16] consists in defining the piston stroke. From run #1006 to #1012, piston stroke varies from 2 to 3 cm with 0.2 cm increment by load adjustment. By increasing the piston stroke, the displacer stroke as well as the phase angle are different for each test. Finally, swept volumes increase with constant operating frequencies. The global model is used to determine the corresponding responses. Figure 14 plots the output power and overall efficiency with respect to the swept volumes. The evolutions of the performances are well represented with a 14.1 % mean departure for powers, 21.8% for efficiencies and 3.2% for swept volumes. Standard deviations are 2.4, 2.4 and 1.1 respectively.

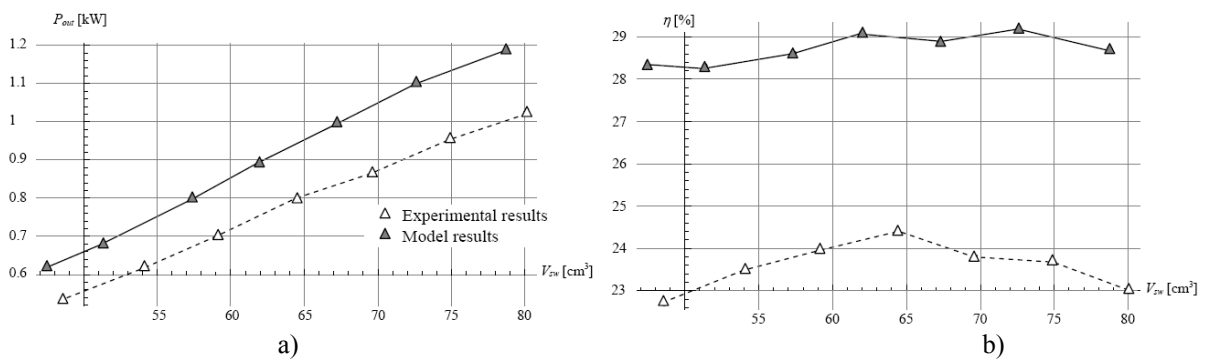

Fig. 14. a) output power, b) overall efficiency as a function of the swept volume for displacer 1.

## 3.3. Influence of piston stroke amplitude for displacer 2

In addition to the evaluation of the effect of the piston stroke amplitude for one configuration (runs #1006 to #1012) another configuration have been chosen (run #1353 to #1359) for model validation. For these tests, the displacer mass is reduced to 381g. Table 8 gives some of the parameters. It is worthy of note that the length of the lighter displacer 2 is less than displacer 1. By doing this, dead volumes of the new engine configuration are more important.

| Heat source $T_H$ | [°C] | 600 | Cold source $T_C$ | [°C] | 25 |
|---|---|---|---|---|---|
| Phase angle $\alpha$ | [deg] | 111.057 | swept volume ratio $\kappa$ | [ - ] | 1.605 |
| Swept volume $V_{sw}$ | [cm$^3$] | 83.7476 | Working fluid | | Helium |
| Mean pressure $p_{mean}$ | [MPa] | 7.038 | Overall efficiency $\eta_{tm}$ [%] | | 21.4 |
| Operating frequency $f$ | [Hz] | 31.1 | Output power [W] | | 987 |
| Wire mesh regenerator 1 | | | regenerator length | [mm] | 64.46 |
| wire diameter $d_w$ | [µm] | 88.9 | porosity of matrix $\P_v$ | [ - ] | 0.759 |
| regenerator hydraulic radius $r_{hR}$ [mm] | | 0.07 | material conductivity | [Wm$^{-1}$K$^{-1}$] | 16.3 |
| Displacer 2 | | | rod diameter | [mm] | 18.085 |
| displacer weight | [g] | 381 | Gas spring volume | [cm$^3$] | 18.8 |

Table 8. Parameter values of Run #1360.

The evolutions of the performances are well represented as can be seen in Fig. 15. The mean departure for powers is 27.9 %, 36.6% for efficiencies and 0.8% for swept volumes. Standard deviations are 7.3, 2.5 and 3.6 respectively.

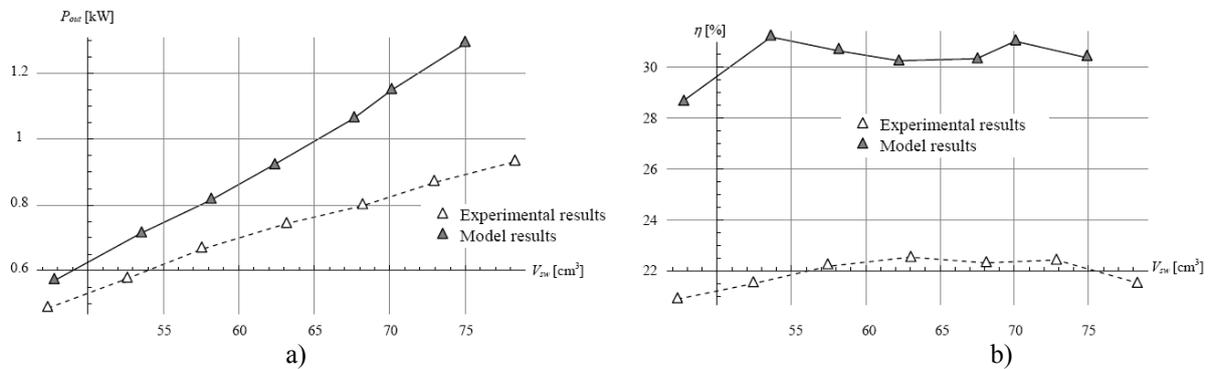

Fig. 15. a) output power, b) overall efficiency as a function of the swept volume for displacer 2.

## 4. Conclusion

Semi analytical thermodynamic and dynamic models of FPSE have been elaborated. The models integrate the regenerator effectiveness, heat exchangers performances as well as conduction losses. Besides a suitable analysis of the pressure drop losses is conducted since the steady state characteristics strongly depends on these effects. Both thermodynamic and dynamic evaluation can be performed in the developed global analysis strategy. The available RE-1000 experimental data have been compared to the model results which show good agreements. Because the geometrical, thermal, fluid and dynamic properties are direct parameters of the global approach, it can be used as a preliminary design tool for FPSE or guidelines for optimization of existing Stirling engines.

## References

[1] Beale W. Stirling cycle type thermal device. US Patent 3552120, 1971.

[2] R.G. Lange, W. P. Caroll, Review of recent advances of radioisotope power systems, *Energy Conversion and Management* **49** (2008), pp. 493-401.

[3] G. R. Schmidt, D. H. Manzella, H. Kamhawi et al., Radioisotope electric propulsion (REP): A near-term approach to nuclear propulsion, *Acta Astronautica* **66** (2010), pp.501-507.

[4] F. De Monte, G. Benvenuto, Reflections on free-piston Stirling engines. Part 1: cycling steady operation, *Journal of Propulsion and Power* **14** 4 (1998), pp.499–508.

[5] F. De Monte, G. Benvenuto, Reflections on free-piston Stirling engines. Part 2: stable operation, *Journal of Propulsion and Power* **14** 4 (1998), pp.509–518.

[6] E. D. Rogdakis, N. A. Bormpilas and I. K. Koniakos, A thermodynamic study for the optimization of stable operation of free piston Stirling engines, *Energy Conversion and Management* **45** 4 (2004), pp. 575-593.

[7] Redlich R. W.; Berchowitz D. M.; Linear dynamics of free-piston Stirling engines, *Proceedings of the Institution of Mechanical Engineers. Part A. Power and process engineering* **199** 4 (1985), pp. 203-213

[8] D.M. Berchowitz, Operational Characteristics of Free-piston Stirling Engines, Proceedings of the 23rd *Intersociety Energy Conversion Engineering*, 1988.

[9] Ulusoy, N. and Mc Caughan, F., Nonlinear analysis of Free Piston Stirling Engine/Alternator System, *Proceedings of 29th Intersociety Energy Conversion Engineering Conference* (1994) pp. 1847-1852.

[10] A.H. Nayfeh, D.T. Mook, Nonlinear oscillations, Wiley, New York, 1979

[11] J.E. Marsden, M. McCracken, The Hopf Bifurcation and its Applications, Applied Mathematical Sciences. Vol. 19, Springer, Berlin, 1976

[12] J. M. Guckenheimer and P. Holmes, Nonlinear Oscillation, Dynamical Systems and Bifurcation of Vector Fields, New York: Springer-Verlag, 1983

[13] A.J. Organ, The Regenerator and the Stirling Engine, *Mechanical Engineering Publications, London*, (1997).

[14] F. Formosa, G. Despesse, Analytical model for Stirling engine design, *Energy Conversion and Management* (2010) doi:10.1016/j.enconman.2010.02.010

[15] W.M. Kays, and A.L. London, Compact Heat Exchangers, *McGraw-Hill* (1964).

[16] J. G. Schreiber, S. M. Geng and G. V. Lorenz, RE-1000 Free-Piston Stirling Engine Sensitivity Test Results, *NASA TM-88846* (1986).

[17] G. Schmidt, The theory of lehmans calorimetric machine, *Z Vereines Deutcher Ingenieure* **15** 1 (1871).

[18] S. K. Andersen, H. Carlsen, P. G. Thomsen, Numerical study on optimal Stirling engine regenerator matrix designs taking into account the effects of matrix temperature oscillations, *Energy Conversion and Management* **47** (2006), pp.894–908.

[19] P.C.T. de Boer, Optimal regenerator performance in Stirling engines, *Int. J. Energy Res*. **33** 9 (2009), pp.813-832.